\documentclass{an}
\usepackage{graphicx}
\usepackage{times}
\usepackage{fancyhdr}
\newcommand{\bec}[1]{\mbox{\boldmath $ #1$}}
\begin{document}

\title{New mechanism of generation of large-scale magnetic fields
 in merging protogalactic and protostellar clouds}

\author{I.\, Rogachevskii \inst{1}, N.\, Kleeorin \inst{1},
 A.D.\, Chernin\inst{2} \and E. Liverts \inst{1}}

 \institute{Department of Mechanical Engineering,
 Ben-Gurion University of Negev, POB 653, 84105 Beer-Sheva, Israel
 \and Sternberg Astronomical Institute, Moscow
 State University, University Prospect 13,
 Moscow 119899, Russia}

\date{Received; accepted; published online}

\abstract{A new mechanism of generation of large-scale magnetic
fields in colliding protogalactic clouds and merging protostellar
clouds is discussed. Interaction of the colliding clouds produces
large-scale shear motions which are superimposed on small-scale
turbulence. Generation of the large-scale magnetic field is due to a
''shear-current" effect (or "vorticity-current" effect), and the
mean vorticity is caused by the large-scale shear motions of
colliding clouds. This effect causes the generation of the mean
magnetic field even in a nonrotating and nonhelical homogeneous
turbulence. There is no quenching of the nonlinear shear-current
effect contrary to the quenching of the nonlinear alpha effect, the
nonlinear turbulent magnetic diffusion, etc. During the nonlinear
growth of the mean magnetic field, the shear-current effect only
changes its sign at some value of the mean magnetic field which
determines the level of the saturated mean magnetic field. Numerical
study shows that the saturated level of the mean magnetic field is
of the order of the equipartition field determined by the turbulent
kinetic energy. The estimated large-scale magnetic field for merging
protogalactic clouds is about several microgauss, and for merging
protostellar clouds is of the order of several tenth of microgauss.
 }

\correspondence{gary@menix.bgu.ac.il}

\maketitle

\section{Introduction}

The generation of magnetic fields in astrophysical objects, e.g.,
galaxies, stars, planets, is one of the outstanding problems of
physics and astrophysics. The initial seed magnetic fields of
galaxies and stars are very weak, and are amplified by the dynamo
process. The generated magnetic field is saturated due to nonlinear
effects.

The origin of seed magnetic fields in the early Universe, e.g., at
phase transitions, is a subject of discussions. However, the merging
of such seed magnetic fields, hardly could produce a substantial
large-scale magnetic fields observed at the present time (Peebles
1980; Zeldovich \& Novikov 1983). The origin of seed magnetic fields
in self-gravitating protogalactic clouds was studied by Birk et al.
(2002) (see also Wiechen et al. 1998). It was suggested that the
seed magnetization of protogalaxies can be provided by relative
shear flows and collisional friction of the ionized and the neutral
components in partially ionized self-gravitating and rotating
protogalactic clouds. Self-consistent plasma-neutral gas simulations
by Birk et al. (2002) have shown that seed magnetic fields $\sim
10^{-14}$ G arise in self-gravitating protogalactic clouds on
spatial scales of 100 pc during $7 \times 10^6$ years.

In this paper we discuss a new mechanism of generation of
large-scale magnetic fields in colliding protogalactic and merging
protostellar clouds. Interaction of the merging clouds causes
large-scale shear motions which are superimposed on small-scale
turbulence. Generation of the large-scale magnetic field is caused
by a ''shear-current" effect or "vorticity-current" effect
(Rogachevskii \& Kleeorin 2003; 2004). The mean vorticity is
produced by the large-scale shear motions of colliding protogalactic
and merging protostellar clouds (Chernin et al. 1991; 1993).

Let us first discuss a scenario of formation the large-scale shear
motions in colliding protogalactic clouds. Jean's process of
gravitational instability and fragmentation can cause a very clumsy
state of cosmic matter at the epoch of galaxy formation. A complex
system of rapidly moving gaseous fragments embedded into rare gas
might appear in some regions of protogalactic matter. Supersonic
contact collisions of these protogalactic clouds might play a role
of an important elementary process in a complex nonlinear dynamics
of protogalactic medium. The supersonic contact non-central
collisions of these protogalactic clouds could lead to their
coalescence, formation of shear motions and transformation of their
initial orbital momentum into the spin momentum of the merged
condensations bound by its condensations (Chernin 1993).

Two-dimensional hydrodynamical models for inelastic non-central
cloud-cloud collisions in the protogalactic medium have been
developed by Chernin (1993). An evolutionary picture of the
collision is as follows. At the first stage of the process the
standard dynamical structure, i.e., two shock fronts and tangential
discontinuity between them arise in the collision zone. Compression
and heating of gas which crosses the shock fronts occurs. The
heating entails intensive radiation emission and considerable energy
loss by the system which promotes gravitational binding of the cloud
material. At the second stage of the process a dense core forms at
the central part of the clump. In the vicinity of the core two kinds
of jets form: "flyaway" jets of the material (which does not undergo
the direct contact collision) and internal jets sliding along the
curved surface of the tangential discontinuity. The flyaway jets are
subsequently torn off, having overcome the gravitational attraction
of the clump whereas the internal jets remain bound in the clump.
When the shock fronts reach the outer boundaries of the clump, the
third stage of the process starts. Shocks are replaced by the
rarefaction waves and overall differential rotation and large-scale
shear motions arise. This structure can be considered as a model of
the protogalactic condensation (Chernin 1993). The formed
large-scale sheared motions are superimposed on small-scale
turbulence.

There are two important characteristics of the protogalactic cloud -
cloud collisions: the mass bound in the resulting clump and the spin
momentum acquired by it. These characteristics depend on the
relative velocity and impact parameter of the collision (Chernin
1993). The parameters of a protogalactic cloud are following: the
mass is $M \leq 10^{10} \, M_\odot$, the radius is $R \sim 10^{23}$
cm, the internal temperature is $T \sim 10^4 $ K, the mean velocity
of the cloud is $V \sim 10^{6} - 10^{7}$ cm/s, where $M_\odot$ is
the solar mass. Some other parameters for the protogalactic clouds
(PGC) are given in Table~1 in Section~3.

An important feature of the dynamics of the interstellar matter is
fairly rapid motions of relatively dense matter fragments
(protostellar clouds) embedded in to rare gas. The origin of
protostellar clouds might be a result of fragmentation of the core
of large molecular clouds. Supersonic and inelastic collisions of
the protostellar clouds can cause merging of the clouds and
formation of a condensation. A non-central collision of the
protostellar clouds can cause conversion of initial orbital momentum
of the clouds in to spin momentum and formation of differential
rotation and shear motions (Chernin 1991). The internal part of the
condensation would have only slow rotation because the initial
matter motions could me almost stopped in the zone of direct cloud
contact. On the other hand, the minor outer part of the merged cloud
matter of the condensation would have very rapid rotation due to the
initial motions of that portions of cloud materials which would not
stop in this zone because they do not undergo any direct cloud
collision (Chernin 1991). This material could keep its motion on
gravitationally bound orbits around the major internal body
condensation. The formed large-scale sheared motions are
superimposed on small-scale interstellar turbulence.

In the supersonic and inelastic collision of the protostellar clouds
an essential part of the initial kinetic energy of the cloud motions
will be lost with the mass lost and also due to dissipation and
subsequent radiative emission. The cooling time scale for the
material compressed in the collision would be less than the time
scale of the hydrodynamic processes. An estimate of basic physical
quantities which characterize the above processes has been made by
Chernin (1991). Thus, e.g., the parameters of a protostellar cloud
are as follows: a mass is $M \leq M_\odot$, the radius is $R \sim
10^{17}$ cm, the internal temperature is $T \sim 10$ K, the mean
velocity of the cloud is $V \sim 10^{5} - 10^{6} $ cm/s. Some other
parameters for the protostellar clouds (PSC) are given in Table~1 in
Section~3.

\section{Generation of large-scale magnetic field due to
the shear-current effect}

Now we discuss generation of large-scale magnetic field due to the
shear-current effect. We suggested that this effect is responsible
for the large-scale magnetic fields in colliding protogalactic
clouds and merging protostellar clouds.

The large-scale magnetic field can be generated in a helical
rotating turbulence due to the $\alpha$ effect. When the rotation is
a nonuniform, the generation of the mean magnetic field is caused by
the $\alpha {\bf \Omega} $ dynamo. For a nonrotating and nonhelical
turbulence the $\alpha$ effect vanishes. However, the large-scale
magnetic field can be generated in a nonrotating and nonhelical
turbulence with an imposed mean velocity shear due to the
shear-current effect (see Rogachevskii \& Kleeorin 2003; 2004). This
effect is associated with the $\bec{\delta} {\bf \times} {\bf J}$
term in the mean electromotive force, where ${\bf J}$ is the mean
electric current. In order to elucidate the physics of the
shear-current effect, we compare the $\alpha$ effect in the $\alpha
{\bf \Omega} $ dynamo with the $\bec{\delta} {\bf \times} {\bf J}$
term caused by the shear-current effect. The $\alpha$ term  in the
mean electromotive force which is responsible for the generation of
the mean magnetic field, reads $ \bec{\cal E}^\alpha \equiv \alpha
{\bf B} \propto - ({\bf \Omega} \cdot {\bf \Lambda}) {\bf B}$ (see,
e.g., Krause \& R\"{a}dler 1980; R\"{a}dler et al. 2003), where $
{\bf \Lambda} = \bec{\nabla} \langle {\bf u}^2 \rangle / \langle
{\bf u}^2 \rangle $ determines the inhomogeneity of the turbulence.
The $\bec{\delta} {\bf \times} {\bf J}$ term in the electromotive
force caused by the shear-current effect is given by $ \bec{\cal
E}^\delta \equiv - \bec{\delta} {\bf \times} (\bec{\nabla} {\bf
\times} {\bf B}) \propto ({\bf W} \cdot \bec{\nabla}) {\bf B} $,
where $\bec{\delta}$ is proportional to the mean vorticity ${\bf W}
= \bec{\nabla} {\bf \times} {\bf U}$ caused by the mean velocity
shear (Rogachevskii \& Kleeorin 2003; 2004).

The mean vorticity ${\bf W}$ in the shear-current dynamo plays a
role of a differential rotation and an inhomogeneity of the mean
magnetic field plays a role of the inhomogeneity of turbulence.
During the generation of the mean magnetic field in both cases (in
the $\alpha {\bf \Omega} $ dynamo and in the shear-current dynamo),
the mean electric current along the original mean magnetic field
arises. The $\alpha$ effect is related to the hydrodynamic helicity
$ \propto ({\bf \Omega} \cdot {\bf \Lambda}) $ in an inhomogeneous
turbulence. The deformations of the magnetic field lines are caused
by upward and downward rotating turbulent eddies in the $\alpha {\bf
\Omega} $ dynamo. Since the turbulence is inhomogeneous (which
breaks a symmetry between the upward and downward eddies), their
total effect on the mean magnetic field does not vanish and it
creates the mean electric current along the original mean magnetic
field.

In a turbulent flow with an imposed mean velocity shear, the
inhomogeneity of the original mean magnetic field breaks a symmetry
between the influence of upward and downward turbulent eddies on the
mean magnetic field. The deformations of the magnetic field lines in
the shear-current dynamo are caused by upward and downward turbulent
eddies which result in the mean electric current along the mean
magnetic field and produce the magnetic dynamo.

\begin{figure}
\centering
\includegraphics[width=8cm]{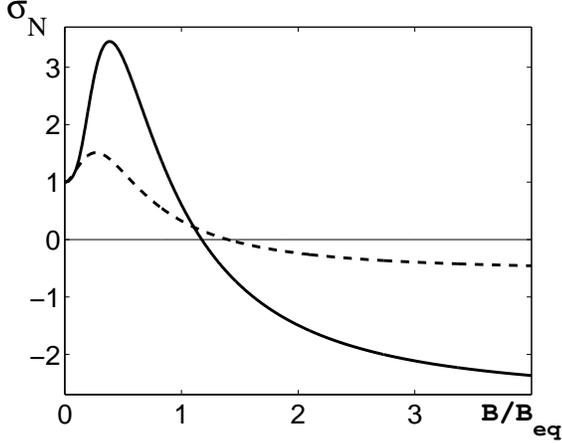}
\caption{\label{Fig1} The dimensionless nonlinear coefficient
$\sigma_{_{N}}(B)$ defining the shear-current effect for different
values of the parameter $\epsilon$: $\, \, \, \epsilon=0$ (solid);
$\epsilon=1$ (dashed).}
\end{figure}

Let us consider for simplicity a homogeneous turbulence with a mean
linear velocity shear, i.e., the mean velocity ${\bf U} = (0, S \,
x, 0)$ and the mean vorticity ${\bf W} = (0,0,S)$. The mean magnetic
field, ${\bf B} = B(x,z) \, {\bf e}_y + (D / S_\ast) \, \bec{\nabla}
{\bf \times} [A(x,z) \, {\bf e}_y]$, is determined by the
dimensionless dynamo equations
\begin{eqnarray}
{\partial A \over \partial t} &=& \sigma_{_{N}}(B) \, \nabla_z B +
\Delta A \;,
\label{F11} \\
{\partial B \over \partial t} &=& - D \, \nabla_z A + \Delta B \;,
\label{F12}
\end{eqnarray}
(Rogachevskii \& Kleeorin 2003; 2004), where $D = (l_0 / L)^2 \,
S_\ast^2 \, \sigma_0 $ is the dynamo number, $S_\ast = S \, L^2 /
\eta_{_{T}}$ is the dimensionless shear number, $\sigma_0 = (4 /
135) \, (1 + 9 \epsilon) $, the parameter $\epsilon$ is the ratio of
the magnetic and kinetic energies in the background turbulence
(i.e., turbulence the with a zero mean magnetic field), $L$ is the
characteristic scale of the mean magnetic field variations,
$\eta_{_{T}}$ is the turbulent magnetic diffusivity,
$\sigma_{_{N}}(B)$ is the function defining nonlinear shear-current
effect which is normalized by $\sigma_0$. We adopted here the
dimensionless form of the mean dynamo equations; in particular,
length is measured in units of $L$, time is measured in units of $
L^{2} / \eta_{_{T}} $ and ${\bf B}$ is measured in units of the
equipartition energy $B_{\rm eq} = \sqrt{4 \pi \rho} \, u_0 $, the
turbulent magnetic diffusion coefficients are measured in units of
the characteristic value of the turbulent magnetic diffusivity
$\eta_{_{T}} = l_0 u_{0} / 3 $, where $u_0$ is the characteristic
turbulent velocity in the maximum scale of turbulent motions $l_0$.
In Eqs.~(\ref{F11}) and (\ref{F12}) we have not taken into account a
quenching of the turbulent magnetic diffusion. This facet is
discussed in details by Rogachevskii and Kleeorin (2004).

The nonlinear function $\sigma_{_{N}}(B)$ defining the shear-current
effect for a weak mean magnetic field $B \ll B_{\rm eq} /4$ is given
by $\sigma_{_{N}}(B) = 1$, and for $B \gg B_{\rm eq} / 4$ it is
given by $\sigma_{_{N}}(B) = - 11 (1 + \epsilon) / 4 (1 + 9
\epsilon)$. The function $\sigma_{_{N}}(B)$ is shown in Fig.~1 for
different values of the parameter $\epsilon$. The nonlinear function
$\sigma_{_{N}}(B)$ changes its sign at some value of the mean
magnetic field $B=B_\ast$. For instance, $B_\ast = 1.2 B_{\rm eq}$
for $\epsilon=0$, and $B_\ast = 1.4 B_{\rm eq}$ for $\epsilon=1$.
The magnitude $B_\ast$ determines the level of the saturated mean
magnetic field during its nonlinear evolution.

Solution of Eqs.~(\ref{F11}) and (\ref{F12}) for the kinematic
problem we seek for in the form $ \propto \exp(\gamma \, t + i K_z
\, z) ,$ where
\begin{eqnarray}
B_y(t,z) &=& B_0 \, \exp(\gamma \, t) \, \cos (K_z z) \;,
\label{M5} \\
B_x(t,z) &=& {l_0 \over L} \, \sqrt{\sigma_0} \, B_0 \, K_z \,
\exp(\gamma \, t) \, \cos (K_z z) \;, \label{M6}
\end{eqnarray}
and we considered for simplicity the case when the mean magnetic
field ${\bf B}$ is independent of $x$. The growth rate of the mean
magnetic field is $\gamma = \sqrt{D} \, K_z - K_z^2$. The wave
vector $K_z$ is measured in units of $L^{-1}$ and the growth rate
$\gamma$ is measured in $ \eta_{_{T}} / L^{2} $. Consider the simple
boundary conditions for a layer of the thickness $2L$ in the $z$
direction, $B(t,|z|=1) = 0$ and $ A'(t,|z|=1) = 0$, i.e., ${\bf
B}(t,|z|=1) = 0$, where $A'$ is the derivative with respect to $z$.
The mean magnetic field is generated when $D > D_{\rm cr} =
\pi^2/4$, which corresponds to $K_z=\pi / 2$. Numerical solution of
Eqs. (\ref{F11}) and (\ref{F12}) with these boundary conditions for
the nonlinear problem is plotted in Fig.~2. In particular, Fig. 2
shows the nonlinear evolution of the mean magnetic field $B(t,z=0)$
due to the shear-current effect for $\epsilon=0$ and different
values of the dynamo number $D$. Here $B(t,z=0)$ is measured in
units of the equipartition energy $B_{\rm eq} = \sqrt{4 \pi \rho} \,
u_0 $.

\begin{figure}
\centering
\includegraphics[width=8cm]{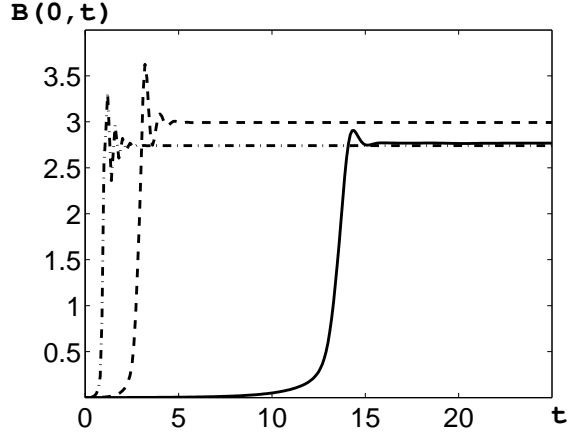}
\caption{\label{Fig2} The nonlinear evolution of the mean magnetic
field $B(t,z=0)$ due to the shear-current effect for $\epsilon=0$
and different values of the dynamo number: $\, \, \, D=\pi^2/4 +1$
(solid); $D=10$ (dashed); $D=40$ (dashed-dotted).}
\end{figure}

The shear-current effect was studied for large hydrodynamic and
magnetic Reynolds numbers using two different methods: the spectral
$\tau$ approximation (the third-order closure procedure) and the
stochastic calculus, i.e., the Feynman-Kac path integral
representation of the solution of the induction equation and
Cameron-Martin-Girsanov theorem (Rogachevskii \& Kleeorin 2003;
2004).  Note that recent studies by R\"{a}dler \& Stepanov (2006)
and R\"{u}diger \& Kichatinov (2006) have not found the dynamo
action in nonrotating and nonhelical shear flows using the second
order correlation approximation (SOCA). This approximation is valid
for small hydrodynamic Reynolds numbers. Indeed, even in a highly
conductivity limit (large magnetic Reynolds numbers) SOCA can be
valid only for small Strouhal numbers, while for large hydrodynamic
Reynolds numbers (fully developed turbulence) the Strouhal number is
unity.

Generation of the large-scale magnetic field in a nonhelical
turbulence with an imposed mean velocity shear was recently
investigated by Brandenburg (2005) and Brandenburg et al. (2005)
using direct numerical simulations. The numerical results are in a
good agreement with the theoretical predictions by Rogachevskii \&
Kleeorin (2004).

\section{Discussion}

In this paper we discussed a new mechanism of generation of the
large-scale magnetic fields in colliding protogalactic and merging
protostellar clouds. Interaction of the merging clouds produces
large-scale shear motions which are superimposed on small-scale
turbulence. The scenario of the mean magnetic field evolution is as
follows. In the kinematic stage, the mean magnetic field grows due
to the shear-current effect from a very small seed magnetic field.
During the nonlinear growth of the mean magnetic field, the
shear-current effect changes its sign at some value $B_\ast$ of the
mean magnetic field. The magnitude ${\bf B}_\ast$ determines the
level of the saturated mean magnetic field. Since the shear-current
effect is not quenched, it might be the only surviving effect, and
this effect can explain the dynamics of large-scale magnetic fields
in astrophysical objects with large-scale sheared motions which are
superimposed on small-scale turbulence.

Note that the magnetic part of the $\alpha$ effect caused by the
magnetic helicity is not zero even in nonhelical turbulence. It is a
purely nonlinear effect. In this study we concentrated on the
nonlinear shear-current effect and do not discuss the effect of
magnetic helicity on the nonlinear saturation of the mean magnetic
field (see, e.g., Kleeorin et al. 2000, 2002; Blackman \&
Brandenburg 2002; Brandenburg \& Subramanian 2005). This is a
subject of a separate ongoing study.

In Table 1 we presented typical parameters of flow and generated
magnetic fields in colliding protogalactic clouds (PGC) and merging
protostellar clouds (PSC). We use the following notations: $\Delta
V$ is the relative mean velocity, $\Delta R$ is the scale of the
mean velocity inhomogeneity, $S = \Delta V / \Delta R$ is the mean
velocity shear, $u_0$ is the characteristic turbulent velocity,
$l_0$ is the maximum scale of turbulent motions, $\tau_0 = l_0 /
u_0$ is the characteristic turbulent time, $\eta_{_{T}}$ is the
turbulent magnetic diffusivity, $t_\eta = (\Delta R)^2 /
\eta_{_{T}}$ is the turbulent diffusion time, $B_{\rm eq} = \sqrt{4
\pi \rho} \, u_0 $ is the equipartition large-scale magnetic field.
Therefore, the estimated saturated large-scale magnetic field for
merging protogalactic clouds is about several microgauss, and for
merging protostellar clouds is of the order of several tenth of
microgauss (see Table 1).

\begin{table}
\label{tab1}
\begin{tabular}{|l|c|c|}
\multicolumn{3}{c}{Table 1}\\
\multicolumn{3}{c}{The parameters of clouds}\\
\hline
                              & PGC     & PSC
 \\
 \hline
   &   & \\
 Mass  & $M \leq 10^{10} \, M_\odot$  & $M \leq M_\odot$ \\
 \hline
   &   & \\
 $R \, $ (cm)  & $R \sim 10^{23}$  & $R \sim 10^{17}$\\
\hline
   &   & \\
 $V \, $ (cm/s)  & $10^{6} - 10^{7}$  & $10^{5} - 10^{6}$
 \\
\hline
   &   & \\
 $\rho \, \,$ (g/cm$^{3}$)        & $10^{-26}$   & $(1-5) \times 10^{-19}$
 \\
 \hline
   &   & \\
 $\Delta V \, $ (cm/s)  & $10^{6} - 10^{7}$  & $10^{5}$
 \\
 \hline
   &   & \\
 $\Delta R \, $ (cm)      & $2 \times 10^{23}$ & $10^{16} - 10^{17}$
 \\
 \hline
   &   & \\
 $S \,$ ($s^{-1}$)        & $(0.5 - 5)  \times 10^{-16}$ & $10^{-12} - 10^{-11}$
 \\
\hline
   &   & \\
 $u_0 \, $ (cm/s)    & $10^{6} - 10^{7}$ & $10^{4}$
 \\
\hline
   &   & \\
 $l_0 \, $ (cm)  & $10^{22}$ & $10^{15} - 10^{16}$
 \\
\hline
   &   & \\
 $\tau_0  \, $ (years)  & $(0.3 - 3) \times 10^{8}$ & $(0.3 - 3) \times 10^{4}$
 \\
\hline
   &   & \\
 $\eta_{_{T}} \,$ (cm$^2$/s)  & $(0.3 - 3) \times 10^{28}$
 & $(0.3 - 3) \times 10^{19}$
 \\
\hline
   &   & \\
 $t_\eta \, $ (years) & $(0.3 - 3) \times 10^{9}$
 & $10^{6} - 10^{7}$
 \\
\hline
   &   & \\
 $B_{\rm eq} \, $ ($\mu$G) & 0.3 - 3  & 10 - 75
 \\
\hline

\end{tabular}
\end{table}

\acknowledgements

This work has benefited from research funding from the European
Community's sixth Framework Programme under RadioNet R113CT 2003
5058187.

\end{document}